\documentstyle[aps,12pt,manuscript]{revtex}
\begin{document}
\preprint{INJE--TP--97--1, hep--th/9701116}
\def\overlay#1#2{\setbox0=\hbox{#1}\setbox1=\hbox to \wd0{\hss #2\hss}#1%
\hskip -2\wd0\copy1}

\title{Negative modes in the four--dimensional stringy wormholes }

\author{Jin Young Kim$^1$, H.W. Lee$^2$, and Y.S. Myung$^2$}
\address{$^1$ Division of Basic Science, Dongseo University, Pusan 616--010, Korea\\
$^2$Department of Physics, Inje University, Kimhae 621--749, Korea} 

\maketitle

\vskip 1.5in

\begin{abstract}
We study the Giddings--Strominger wormholes in string theories.
We found a non--singular wormhole solution and analyzed the
perturbation around this wormhole solution.  We have used the bilinear
action to obtain Schr\"odinger--type equation for perturbation
fields assuming a linear relation between the perturbation fields.
With this analysis, we found an infinite number of negative modes
among $O(4)$--symmetric fluctuations about the non--singular wormhole
background.
\end{abstract}

\newpage
Euclidean wormholes--solutions to the euclidean 
Einstein equations that connect two asymptotically
flat regions--are considered as saddle points of the 
functional integral and
are very important for semiclassical calculations of 
transition probabilities of
 topological change  in quantum gravity.
There are many kinds of euclidean wormhole solutions.
In four--dimensions the following matters which support the 
throat of the wormhole were adopted:
axion fields\cite{Gid1}, scalar fields\cite{Lee}, SU(2) 
Yang--Mills fields\cite{Hos}.
 Higher--dimensional wormhole solutions
were obtained \cite{Yos,Mye} and higher--derivative correction 
to the Einstein--Hilbert
 action was considered\cite{Fuk}. Recently, we found the 
D--wormhole solution in type IIB 
superstring theory\cite{Kim}. However, it turned out to be a 
ten--dimensional singular 
wormhole with infinite euclidean action density.

On the other hand, we are interested in the contribution of 
wormhole configurations to
 the euclidean functional integral
for the forward ``flat space $\to$ flat space '' amplitude.
Rubakov and Shvedov\cite{Rub} decided semiclassically 
whether Giddings--Strominger wormhole  
makes real or complex contributions into the functional integral 
in four--dimensional curved space . On the analogy of
the analysis of instantons/bounces in the quantum field theory, 
it is found that
the wormhole contribution
is imaginary since there exists one negative mode ($\omega^2 =-4$)
among fluctuations  around the classical euclidean solution.
 This means that the classical solution 
with one negative mode is not stable against the fluctuations and thus
belongs to the bounce.

In this paper, we study  the Giddings--Strominger wormholes in string 
theories\cite{Gid2}.
Hereafter we wish to call these as stringy wormholes to distinguish 
the previous 
Giddings--Strominger wormhole.
We have both the singular wormhole  as well as the non--singular one.
It is carried out the analysis of 
$O(4)$--symmetric fluctuations
 about the non--singular wormhole background.
We use the bilinear
action to obtain Schr\"odinger--type equation for perturbation
fields assuming a linear relation between perturbation fields.
With this analysis, we find an infinite number of negative modes
among $O(4)$--symmetric fluctuations about the non--singular wormhole
background.

Our analysis is similiar to the stability analysis of the black 
holes\cite{Chan},
which is classical solution in curved spacetime with the Minkowski
signature.  One easy way of understanding  a black hole  is to find 
out how it reacts to external perturbations. 
We can visualize the 
black hole  as presenting an effective potential barrier (or well) to the 
on--coming waves. 
 As a  compact criterion for the  black hole case, 
 it is unstable if there
exists a potential well to the on--coming waves.
 This is so because  the Schr\"odinger--type equation 
with the potential well always allows the bound states as well as 
scattering states. 
The former shows up as an imaginary frequency mode ($\omega^2 <0$), 
leading to an exponentially growing mode with time.
If one finds any exponentially growing perturbation, the 
black hole turns out to be unstable.

Our starting action is the NS--NS sector of ten--dimensional string 
theory\cite{Gid2},
\begin{equation}
S_{10} = \int d^{10} x \sqrt{g_{10}}~ e^\phi 
  \big[ -  R - (\nabla \phi)^2 + H^2  \big],
\label{action}
\end{equation}
where $\phi$ is the dilaton and $H=dB$ with a NS--NS two--form $B$.
Here we do not consider the R--R sector for simplicity\cite{Wit}.
The ten--dimensional theory can be reduced to four--dimensional one 
by the compactification
on a six--dimensional Calabi--Yau manifold. This is realized 
($M^{10}\to M^4 \times M^6$)
 by giving the following
vacuum expectation values: 
\begin{eqnarray}
&& \bar g_{MN} =
 \left(  \begin{array}{cc}  \tilde g_{\mu\nu}(x) & 0   \\
                             0 & e^{D(x) / \sqrt 3} g_{mn}(y)  \end{array}   \right),
                           \nonumber   \\
&& \bar B_{\mu\nu} = B_{\mu\nu} (x),  \\
&&  \bar B_{mn} = (1/6) a(x) b_{mn} (y),    \nonumber \\
&& \bar \phi = \phi(x).         \nonumber  
\end{eqnarray}
and the rest of fields will be taken to zero.
Here  $\mu,\nu, \cdots (m,n,\cdots)$ denote four (six)-dimensional 
indices, and $x(y)$
represent four (six)-dimensional coordinates.
The field equations for the graviton, dilaton, and two--form field are 
satisfied if the 
internal manifold ($M^6$) is Calabi--Yau (Ricci--flat and K\"ahler) and 
the equations of motion obtained from  the  four--dimensional 
effective action
\begin{equation}
S_{4} = \int d^4 x \sqrt{g}
  \big[ -  R +{1 \over 2} (\nabla D)^2 +{1 \over 2} e^{ - {2 \over \sqrt 3} D} (\nabla a)^2 
      +{1 \over 2} (\nabla \Delta)^2 + e^{2 \Delta} H^2  \big],
\label{action4}
\end{equation}
where
\begin{equation}
\Delta \equiv \phi + \sqrt 3 D,  ~~~~~g_{\mu\nu} = e^\Delta \tilde g_{\mu\nu},
\label{delta}
\end{equation}
are satisfied.

Now let us find stringy wormhole solution by considering either 
$a$ or $B_{\mu\nu}$ as
matter which supports the throat of the wormhole. Here we
confine our main interest to the first case (the non--singular
wormhole). The latter case leads to the singular wormhole.  The
non--singular case is realized when $H =\Delta =0$. The action is
given by
\begin{equation}
S = \int d^4 x \sqrt{g} 
  \big[ -  R +{1 \over 2} (\nabla D)^2 +
         {1 \over 2} e^{ - {2 \over \sqrt 3} D} (\nabla a)^2  \big].
\label{action0}
\end{equation}
One can consider $a(x)$ as the source of the wormhole. 
We thus take the Noether current $J_\mu =e^{-(2/\sqrt 3) D}\partial_\mu a$ and require 
its conservation
\begin{equation}
\partial_\mu (\sqrt {g} J^\mu) =0.
\label{noether}
\end{equation}
Therefore we have to perform the functional integration over 
conserved current densities.   
We introduce the general $O(4)$--symmetric euclidean metric as

\begin{equation}
 ds^2= N^2(\rho) d\rho^2 + R^2(\rho) d\Omega^2_3
\label{proper}
\end{equation}
with two scale factors $(N, R)$.
The $O(4)$--symmetric current density has one non--zero component 
($J^0(\rho))$ and its
conservation in (\ref{noether}) means that $\sqrt g J^0$ is a
constant. This constant is related to the 
global charge $Q$ of the wormhole $(Q / Vol(S^3))$. Thus one finds

\begin{equation}
 J^0 = { Q \over 2 \pi^2} {1\over N R^3}.
\label{j0}
\end{equation} 
The action (\ref{action0}) can be rewritten as 
\begin{equation}
S =  6 \int d^4 x   
  \big[ - {R R^{\prime 2} \over N} - N R + {1 \over 12} {R^3 \over N} D^{\prime 2}
        +{ Q^2 \over 48 \pi^4} { N \over R^3} e^{ {2 \over \sqrt 3} D }
  \big],  
\label{action2}
\end{equation}
where the prime means the derivative with respect to $\rho$.
From the above action, the equations of motion are 
\begin{equation}
     {R R^{\prime 2} \over N^2} -  R - {1 \over 12} {R^3 \over N^2} D^{\prime 2}
        +{ Q^2 \over 48 \pi^4} { 1 \over R^3} e^{ {2 \over \sqrt 3} D } = 0,   
\label{eom1}
\end{equation}
\begin{equation}
     - { R^{\prime 2} \over N} + 2 \big ( { R R^\prime \over N} \big )^\prime
     - N  + {1 \over 4} {R^2 \over N} D^{\prime 2}
     - { Q^2 \over 16 \pi^4} { N \over R^4} e^{ {2 \over \sqrt 3} D } = 0,
\label{eom2}
\end{equation}
\begin{equation}
     - { 1 \over 6} \big ( { R^3 D^\prime \over N} \big )^\prime
     + { Q^2 \over 24 \sqrt{3}  \pi^4} { 1 \over N R^3} e^{ {2 \over \sqrt 3} D } = 0.
\label{eom3}
\end{equation}
For $N=1$ gauge, (\ref{eom1}) and (\ref{eom3}) are reduced to
\begin{equation}
      R^{\prime 2} = 1 + {1 \over 12} R^2  D^{\prime 2}
        - { Q^2 \over 48 \pi^4} { 1 \over R^4} e^{ {2 \over \sqrt 3} D },
\label{eom4}
\end{equation}
\begin{equation}
     \big ( R^3 D^\prime \big )^\prime
     = { Q^2 \over 4 \sqrt{3}  \pi^4} { 1 \over R^3} e^{ {2 \over \sqrt 3} D }.
\label{eom5}
\end{equation}
From (\ref{eom5}), one finds the dilaton equation
\begin{equation}
      R^6 D^{\prime 2}
     = { Q^2 \over 4 \pi^4} e^{ {2 \over \sqrt 3} D }
      - { Q^2 \over 4 \pi^4} e^{ {2 \over \sqrt 3} D_0 },
\label{eom6}
\end{equation}
where the integration constant is chosen so that $D$ has vanishing 
derivative at the wormhole
neck ($\rho=0$).
Substituting this into (\ref{eom4}), one obtains 
\begin{equation}
      R^{\prime 2} = 1 - {R_0^4 \over R^4},~~~~
      R_0^4= { Q^2 \over 48 \pi^4} e^{ {2 \over \sqrt 3} D_0}.
\label{classic}
\end{equation}
Here $R_0=R(\rho=0)$ corresponds to the radius of wormhole neck ($R^\prime =0$).
Equation (\ref{eom2}) is satisfied with (\ref{eom6}) and
(\ref{classic}) and thus is a redundant one. The resulting solution
(stringy wormhole) to (\ref{classic}) has the asymptotic behavior
$R(\rho) \to \pm \rho$ as $\rho \to \pm \infty$, corresponding to two
asymptotically flat regions and has minimum at $R_0$. Further
(\ref{eom6}) is solved to obtain
\begin{equation}
e^{ - {2 \over \sqrt 3} D }= { Q^2 \over 48 \pi^4} {1 \over R^4},
\label{d-r}
\end{equation}
which will prove very useful for the computation of the perturbed 
action on later.
Note that $D$ is non--singular for finite $\rho$ and thus the 
integrand of the action is finite too.
For an explicit calculation, we  wish to solve the differential 
equation (\ref{classic})
 by numerical analysis.
  We introduce the rescalings
 $(\rho/\rho_0, R/R_0, D/D_0)$ with $\rho_0=R_0$. 
 The resulting solution 
is shown in Fig.1. 
  Far from the wormhole throat $(\rho/R_0 > 1)$, 
one can ignore the effect of gravity
and the euclidean space becomes flat $(R \sim \rho)$.  Here one 
can find the wormhole neck
($ R^\prime=0$) near $\rho = 0$.  
Now let us substitute the results of $R(\rho)/R_0$ in Fig.1 into
(\ref{d-r}). Then one obtains the behavior of the wormhole dilaton
($D(\rho)$). As is shown in Fig.2, $D$ does not have any singular
point.

Let us now consider $O(4)$--symmetric fluctuations about the 
non--singular wormhole solution.
In general, the interpretation of the 
wormhole
 depends on whether or not there are negative
modes around the solution. If one finds odd number of negative modes,
 the solution corresponds to
a bounce and describes the nucleation and  growth of wormhole in 
the Minkowski spacetime.  
If there are even number of negative modes, the path integral would be
real and classical solution would resemble an instanton rather than a
bounce. If there is no negative mode, the solution is called an
instanton and describes the tunneling and  mixing of two states of the
same energy. The small fluctuations are given by
\begin{equation}
R(\rho)= R_c(\rho) + r(\rho),~~~ N(\rho) = 1 + n(\rho),~~~~ 
D(\rho) = D_c(\rho) + d(\rho),
\label{pertub}
\end{equation}
where $R_c, D_c$ represent the classical wormhole background.
Substituting these into (9) and then take only the bilinear parts 
in $(r, n, d)$ of the action. This is because from this part 
one can derive the linearized equations which are essential for the 
fluctuation study.
Here we choose the $n(\rho)=0$ gauge,
since the quadratic action is invariant under the $O(4)$--general 
coordinate transformations.
The bilinear action is then given by

\begin{eqnarray}
S_{bil} =  12 \pi^2 \int d \rho   
  \Big[ &-& R_c r^{\prime 2} -2 R_c^\prime r r^\prime
      + {1 \over 12} ( R^3 d^{\prime 2} + 6 R^2_c D_c^\prime r d^\prime + 
         3 R_c D_c^{\prime 2} r^2) \nonumber      \\
        &+&{ Q^2 \over 48 \pi^4} { 1 \over R_c^3} e^{ {2 \over \sqrt 3} D }
      ( 6 {r^2 \over R_c^2} - 2 \sqrt{3} {d r \over R_c} + {2 \over 3} d^2) 
 \Big].
\label{sbil}
\end{eqnarray}
After some calculation, (\ref{sbil}) can be rewritten as 
\begin{eqnarray}
S_{bil}  =12 \pi^2 \int d \rho   
  \Big[ &-& R_c r^{\prime 2} +({9 \over R_c} - {R_0^4 \over R_c^5}) r^2 
      + {R_c^3 \over 12} d^{\prime 2}              \nonumber      \\
        &+&{ 1 \over 2} R_c^2 D_c^\prime r d^\prime 
         - 2 \sqrt{3} d r  + {2 \over 3} R_c d^2  \Big]
\label{rsbil}
\end{eqnarray}
with the boundary terms which are not relevant for our study.
One confronts with difficulty in dealing with (\ref{rsbil}).   This is
because of the presence of
$r$-$d$ coupling terms. Actually one has to find the new canonical 
variables that diagonalize
the action (\ref{rsbil}).   However, thanks to the relation
(\ref{d-r}), one has the relation between 
$r$ and $d$.  Linearizing (\ref{d-r}) leads to $ d = 2 \sqrt{3}  r /
R_c$ and inserting this
into (\ref{rsbil}), we obtain $S_{bil} = 0$ which leads to a trivial
case.  In order to avoid this trivial case,
 we assume the relation as  
\begin{equation}
d = 2 \sqrt{3} \alpha {r \over R_c}
\label{pd-r}
\end{equation}
by introducing $\alpha$ as the parameter.  This means that $d$ is not 
an independent variable.  The above is the simplest
assumption which is appropriate in the spirit of linear
perturbation.  Otherwise, the analysis becomes very difficult.  
Using (\ref{pd-r}), we find the desirable bilinear form
\begin{equation}
S_{bil} =  12 \pi^2 (\alpha^2 -1)  \int d \rho   
  \Big[  R_c r^{\prime 2} + ({R_0^4 \over R_c^5} + 
   {\alpha -1 \over \alpha + 1} {9 \over R_c}) r^2 
\label{reduce}
\Big].
\end{equation}
One can easily check that $S_{bil}=0$ for $\alpha =1$.  
Since the bilinear from (\ref{reduce}) is positive definite for
$\alpha^2 > 1$, there is no negative modes in this region.  Thus the
range of the parameter should be confined to $\alpha^2 <1 $. But for
$\alpha^2 < 1$, the action is unbounded from below, because of the
negative sign of the kinetic term. In this case,  we need the GHP
rotation\cite{Gib} for scale factor ($r \to ir$). Taking the variation
of the action (\ref{reduce}) with respect to $r$, on gets the
Schr\"odinger--type equation
\begin{equation}   
  R_c \Big [ - (R_c r^\prime )^\prime 
+ ({R_0^4 \over R_c^5} + {\alpha -1 \over \alpha + 1} {9 \over R_c}) r  \Big] = \omega^2 r.
\label{sch}
\end{equation}
Here we choose a prefactor $R_c$ on the left--hand side in such a way
 that the above equation can be solved explicitly.  
From now on we are interested in negative mode $\omega^2 =-|\omega|^2$. 
For simplicity we set $R_0=1$.
Introducing a new variable $y = R_c^{-4}$ and rewriting $r = R_c^{-p} \psi (y)$, 
 (\ref{sch}) is reduced to the form of a hypergeometric equation 
\begin{equation}   
y (1 -y) {d^2 \psi \over dy^2} 
+\big \{ (1 + {p \over 2}) - ({3 \over 2} + {p \over 2}) y \big \} {d \psi \over d y}
- {1 \over 16} (p+1)^2 \psi = 0,
\label{hyper}
\end{equation}
with 
\begin{equation}
 p^2 = |\omega|^2 + 9 {\alpha -1 \over \alpha +1}.
\label{p2}
\end{equation}
Note that the variables $y$ and $\rho$ are not in one to one correspondence.
This can be cured by requiring that $r(\rho)$ is either symmetric or antisymmetric
with respect to $\rho$. One finds that $y$ and $1-y$ are symmetric,
while $\sqrt{1-y}$ is antisymmetric in $\rho$. The symmetric  solution
of (\ref{hyper}) is 

\begin{equation}   
\psi (y) =  C_1 F({p+1 \over 4}, {p+1 \over 4}; 1 + {p \over 2}; y)
         +  C_2 F({p+1 \over 4}, {p+1 \over 4};  {1 \over 2}; 1-y),
\label{psi}
\end{equation}
where $C_1$ and $C_2$ are arbitrary constants. 
Further one requires the small perturbation such that $r(\rho)$ be finite
at both ends ( $\rho \to \infty (y \to 0)$ and $\rho \to 0 (y \to
1)$ ). The last term in (\ref{psi}) behaves $y^{-( p + 1)^2/8}$ as $y
\to 0$, and this  in turn gives us 
$r_{y \to 0} \to y^{-( p^2 + 1)/8}$. This diverges as $y \to 0$ and thus 
we set $C_2=0$.
On the other hand, the first term in (\ref{psi}) is finite at both
ends. We also impose that the eigenfunction $r$ is square integrable
with the weight  $d\rho/R_c(\rho)$.
This is compatible with the choice of the prefactor $R_c$ in
(\ref{sch}). The integrability condition is realized as
\begin{equation}
\int^\infty_0 { r^2 d\rho \over R_c} = \int^1_0 \psi^2 y^{p/2 -1} (1-y)^{1/2 -1}dy
< M \int^1_0  y^{p/2 -1} (1-y)^{1/2 -1}dy = M B(p/2, 1/2),
\label{bound}
\end{equation}
where $M$ is the maximum value of $\psi^2$ in [0,1] and 
$B(p/2,1/2)= \Gamma(p/2) \Gamma(1/2) / \Gamma(p/2+1/2)$ is the beta--function.
The condition for finite $B(p/2, 1/2)$ requires that $p$ should be positive.
From (\ref{p2}), this condition is satisfied if  
 $|\omega|^2> -9 (\alpha -1)/(\alpha +1)$ for $\alpha^2<1$.
Hence one can always find negative modes for all positive $p$.

We perform the analysis of $O(4)$--symmetric fluctuations on the
stringy  wormhole background
with the gauge $n(\rho)=0, r(\rho) \not=0$.
Instead of diagonalizing the quardratic action, we choose 
 the relation $d= 2 \sqrt 3
\alpha r / R_c$ which is inspired by (\ref{d-r}).  
Rubakov and Shvedov\cite{Rub} reported that there exists only one 
negative mode $r^{(-)}
(\rho) = 1 / R^2_c(\rho)$ with $\omega^2= -4$ for pure gravity case.
The existence of one negative mode implies that the wormhole
contribution into the functional integral is imaginary, which
corresponds to the instability of the parent universe against the
emission of a baby universe.  In our case $r(\rho)= R^{-p}_c$ with
$p=-1$ satisfies (\ref{sch}) over the entire region. But this solution
is not a small perturbation and thus we discard it. On the other hand,
we find a continuous spectrum of negative modes for positive $p$  by
requiring both symmetric property and integrability.  This diffrence
comes from the fact that the last term of (\ref{hyper}) is different
from (8) in Ref. \cite{Rub}.  It has been shown by Coleman\cite{Col}
that the bounce interpretation of a classical solution requires
exactly one negative mode.  In general, the reality and
imaginarity of the path integral depends on the sign of the
determinant of fluctuatuons.  The essential property is whether the
number of negative modes is odd or even.  The odd case belongs to the
bounce, while the even case is related to the instanton.  Here we
obtain the continuous spectrum of negative modes.  The existence of 
an infinite number
of negative modes leads to diffrent problems.  Lavrelashvili,
Rubakov and Tinyakov(LRT)\cite{Lav} pointed out that an infinite number
of negative modes may appear around the bounce.  But Tanaka and
Sasaki\cite{Tan} argued that the above LRT claim is an artifact due
to their inadequate choice of gauge (LRT gauge), which was inevitably
implied by the Lagrangian formalism.
For the LRT gauge of $n(\rho) \not=0, r(\rho)=0$ in\cite{Tan}, 
one can obtain the bilinear action from (\ref{action2}).
One has to use the constraint equation (\ref{eom1}) to eliminate the
$n(\rho)$--terms.  Unfortunately,
we cannot get the relation between $n(\rho)$ and $d(\rho)$ by
linearizing (\ref{eom1}).  Further the corresponding
action turns out to be trivial. 

In our case, we choose the gauge of $n(\rho) =0, r(\rho) \not=0$.
Under this gauge, one has to perform the Hamiltonian analysis arisen
from Ref.\cite{Tan}.  At this stage, it is not clear to conclude
whether the stringy wormhole is a bounce or an instanton. 

\acknowledgments

This work was supported in part by the Basic Science Research Institute 
Program, Korean Ministry of Education, Project 
NOs. BSRI--96--2413, BSRI--96--2441
and by Inje Research and Scholarship Foundation.

\figure{Fig. 1: $R/R_0$ as a function of $\rho / R_0$.  The solid, 
dotted and dashed
lines correspond to wormhole scale factor 
($R/R_0$), $R(\rho)/R_0 \approx1.274$, and 
$R/R_0= \rho$. 
The singular point ($\rho_{sg}$) is determined as a solution to
 $R(\rho_{sg})/R_0 =1/ \sqrt {\cos (\pi/2 \sqrt 3)} \approx 1.274$.

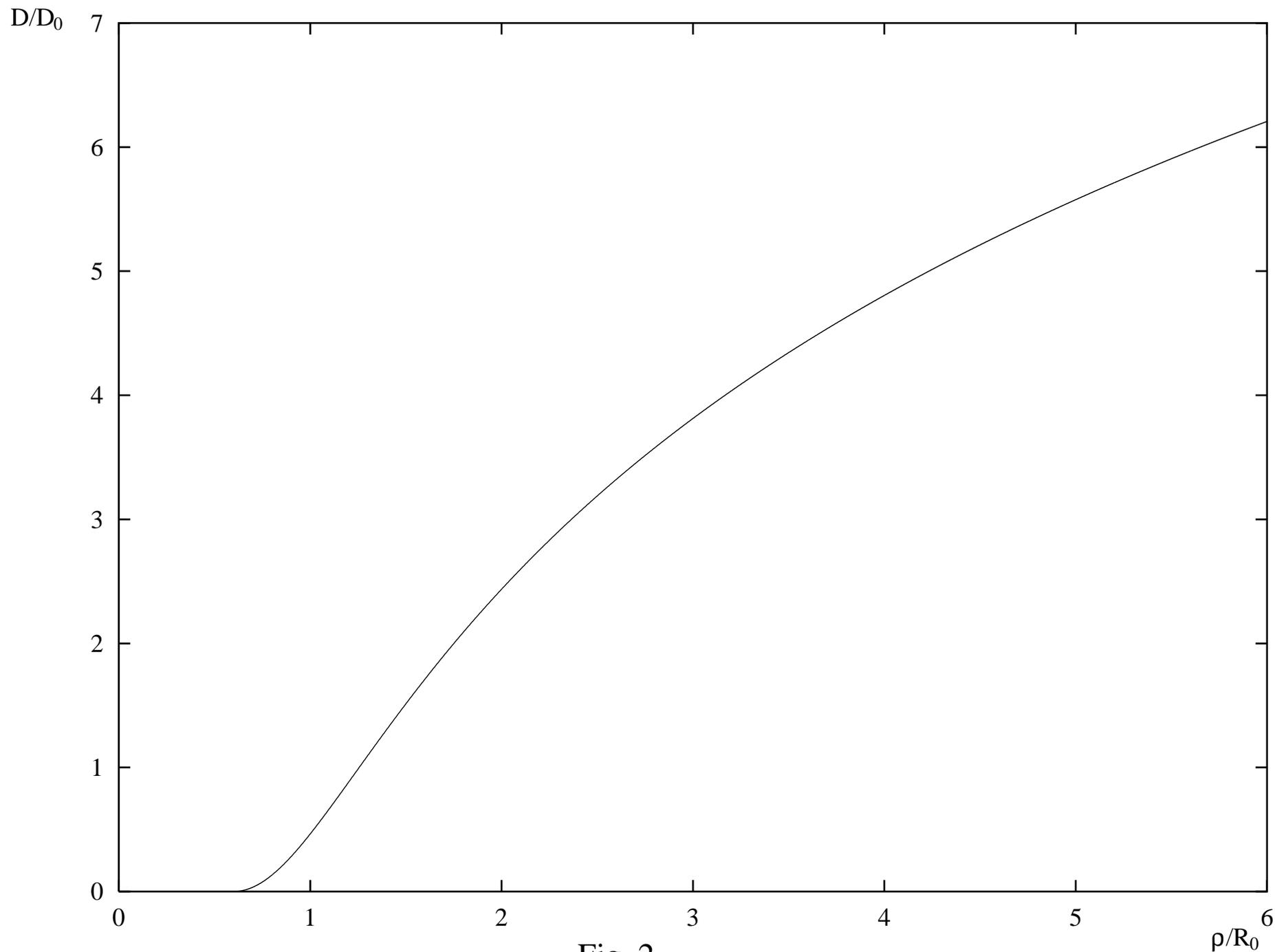
Fig.2: $D/D_0$ as a function of $\rho / R_0$.  
No singular point  is found. 
}


\newpage

\begin{references}
\bibitem{Gid1} Giddings and A. Strominger, Nucl. Phys. {\bf B306} (1988) 890.
\bibitem{Lee}  K. Lee,  Phys. Rev. Lett. {\bf 61} (1988) 263;
                B. Grinstein, Nucl. Phys. {\bf B321} (1989) 439.
               L. F. Abbott and M. B. Wise, Nucl. Phys. {\bf B325} (1989) 687;
               S. Coleman and K. Lee, Nucl. Phys. {\bf B329} (1990) 387;
               B. Grinstein, Nucl. Phys. {\bf B321} (1989) 439.
\bibitem{Hos}    A. Hosoya and W. Ogura,  Phys. Lett. {\bf B 225} (1989) 117.
\bibitem{Yos}   K. Yosida, S. Hienzaki, and K. Shiraishi, Phys. Rev. {\bf 42} (1990) 1973.
\bibitem{Mye}    R. C. Myers, Phys. Rev. {\bf D38} (1988) 1327.             
\bibitem{Fuk}   H. Fukutaka, K. Ghoroku, and K. Tanaka,  Phys. Lett. {\bf B 222} (1989) 191.   
\bibitem{Kim}  J. Y. Kim, H. W. Lee, and Y. S. Myung, hep--th/9612249
(Phys. Lett. B, to be published).
\bibitem{Rub}  V. A. Rubakov and O. Shvedov,  Phys. Lett. {\bf B383} (1996) 258.            
\bibitem{Gid2}  G. W.  Giddings and A. Strominger, Phys. Lett. {\bf B230} (1989) 46;
                 S. J. Rey, Phys. Rev. {\bf D43} (1991) 526. 
\bibitem{Chan} S. Chandrasekhar,  {\it The Mathematical Theory of Black Hole}
                 (Oxford Univ. Press, New York, 1983);
               O. J. Kwon, Y. D. Kim, Y. S. Myung, B. H. Cho and Y. J. Park,
               Phys. Rev. {\bf D34} (1986) 333 ; Int. J. Mod. Phys. {\bf A1} (1986) 709.
\bibitem{Wit} E. Witten, Nucl. Phys. {\bf B443} (1995) 85;
              D. Polyakov, Nucl. Phys. {\bf B468} (1996) 155; 
               A. A. Tseytlin, Class. Quant. Grav. {\bf 13} (1996) L81;
               V. Balasubramanian and F. Larsen, hep--th/9610077.
\bibitem{Gib} G. W. Gibbons, S. W. Hawking, and M. J. Perry, Nucl. Phys.{\bf B 138} (1978) 141.
\bibitem{Ruba}  V. A. Rubakov and O. Shvedov, gr--qc/9608065. 
\bibitem{Col} S. Coleman, Nucl. Phys. {\bf B298} (1988) 178.
\bibitem{Lav} G. V. Lavrelashvili, V. A. Rubakov and P. G. Tinyakov, 
              Phys. Lett. {\bf B161} (1985) 280.
\bibitem{Tan} T. Tanaka and M.
Sasaki, Prog. Theor. Phys. {\bf 88}(1992) 503.    
\end{references}
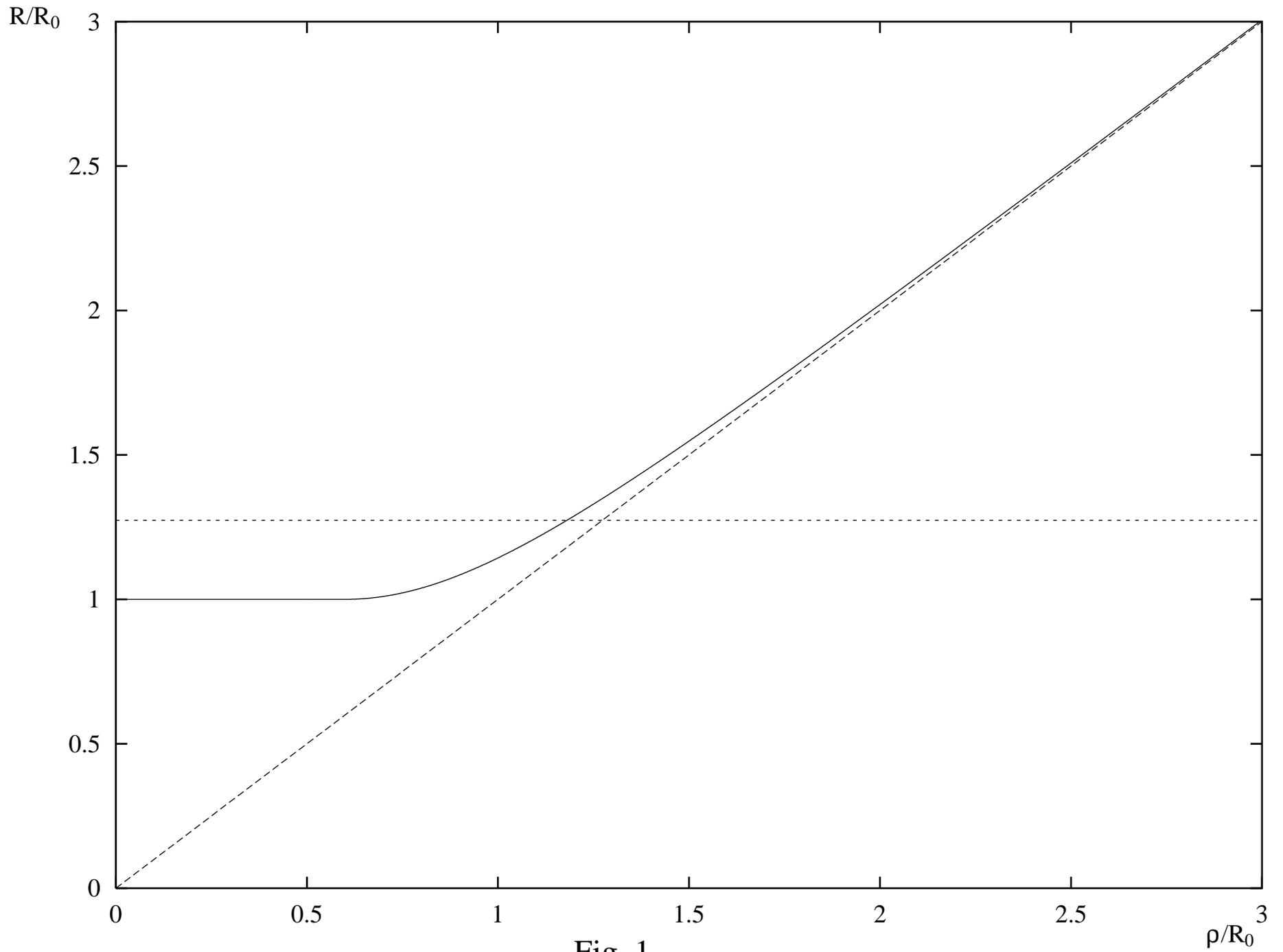
\end{document}